# Execution Timing Control for Deterministic Task Offloading in the IoT-Edge-Cloud Continuum

Keyvan Aghababaiyan, Baldomero Coll-Perales, Javier Gozalvez
*Networked Systems Lab, Universidad Miguel Hernandez de Elche, Elche (Alicante), Spain*
kaghababaiyan@umh.es, bcoll@umh.es, j.gozalvez@umh.es

***Abstract*—Latency-critical IoT applications, such as autonomous mobility and industrial automation, require deterministic guarantees to ensure that tasks are completed within strict deadlines. The 6G-enabled IoT–edge–cloud continuum can support such requirements by leveraging communication, computation and intelligence resources across devices, edge, and cloud infrastructures. However, existing task offloading strategies mainly focus on selecting where tasks are executed and typically assume immediate processing upon task arrival. This leads to transient congestion when multiple tasks coincide in time and results in inefficient resource utilization under dynamic workloads. This paper addresses these limitations by introducing an execution timing control strategy for deterministic task offloading that jointly determines where tasks are executed and when their execution starts, while guaranteeing deadline compliance. The key idea is to exploit the latency budget of tasks to control their execution timing, enabling a more balanced distribution of workload over time and reducing peak congestion across the continuum. Evaluation results show that, compared to existing benchmarks, the proposed approach achieves up to 70% higher satisfaction ratio, reduces the communication resources usage by 40%, lowers peak computing resource utilization by 15%, and decreases average execution time by up to 77%.**

## I. INTRODUCTION

Many applications in critical IoT verticals, including autonomous mobility and industrial automation, require deterministic execution capabilities to guarantee timely decision-making within bounded latency deadlines. Failure to satisfy these constraints may result in unsafe situations, operational disruptions, or system instability. Supporting such verticals with future cellular networks requires system designs capable of guaranteeing deterministic end-to-end service levels [1]. The 6G-driven IoT–edge–cloud continuum has emerged as a key enabler for supporting vertical applications by integrating computing, connectivity, and intelligence resources across IoT devices, edge nodes, and cloud infrastructures into a unified environment. This integration enables task offloading to any node in the continuum, allowing applications to exploit the heterogeneous capabilities available across the continuum. However, fully leveraging this flexibility requires efficient and coordinated resource management across the continuum. Offloading a task to the edge or cloud can reduce processing time but may introduce additional communication latency, whereas local execution avoids communication latency but is constrained by limited device capabilities. Therefore, effective task offloading strategies must jointly consider task requirements, communication and computation latencies, and the availability of resources across the continuum [2].

Most existing task offloading works focus on selecting the computing unit where each task should be processed. For example, [3]–[5] select the computing unit with the objective of minimizing individual task latency. [6] studies latency-aware task execution in mobile edge computing, where tasks are processed locally or offloaded to nearby edge servers to reduce latency. [5] extends this approach to multi-layer IoT–edge–cloud systems, aiming to minimize task execution latency. While these approaches can reduce latency for many tasks, they may lead to transient congestion and system-level performance degradation when multiple concurrent tasks overload highly demanded computing units. As a result, decisions that are optimal for individual tasks may not be optimal at the system level. The study in [7] shows that a deterministic task offloading strategy can achieve better system-level performance by prioritizing the maximization of the number of tasks completed within their deadlines, rather than focusing on minimizing individual task latency. This deterministic strategy leverages tasks' latency budgets to balance the workload across the continuum by offloading tasks to nodes that can complete them within their deadlines, even if their execution latency is not minimized. As a result, it mitigates congestion, improves resilience to workload variations, and increases the number of tasks executed in time. However, the deterministic strategy in [7] only determines where tasks are executed and does not exploit latency budgets to control when their execution should start. While the concept of execution timing control has been studied in the context of static scheduling for single-node embedded systems [8], its potential for managing the continuum remains unexplored. In this paper, we study such potential by integrating execution timing control into deterministic task offloading strategies in the continuum.

This paper advances the state of the art by introducing execution timing control into deterministic task offloading.

The work of K. Aghababaiyan is supported in part by the European Union under the 2024 MSCA Postdoctoral Fellowship program (project no. 101207598). This work has been also partially funded by MCIN/AEI/10.13039/501100011033 (PID2023-150308OB-I00), UMH *and Generalitat Valenciana* (CIAICO/2024/167).

The proposed approach jointly determines where tasks are executed and when their execution begins, enabling workload distribution not only across nodes in the continuum but also over time. By exploiting the latency budget of tasks to control their execution timing, the proposed strategy improves resource utilization across the continuum, reduces congestion, and increases the number of tasks completed within their latency deadlines. Our evaluation shows that incorporating execution timing control significantly enhances the ability to support deterministic service levels while reducing latency through more efficient utilization of resources across the continuum and over time.

## II. SYSTEM MODELING

We consider an IoT-edge-cloud continuum where multiple IoT devices can connect to edge computing nodes and a centralized cloud platform through a wide-area cellular network. We assume a scenario where tasks originating from the IoT devices can either be executed locally at the devices or offloaded to the edge or cloud, constrained by the available resources in the continuum and the tasks' execution deadlines. Specifically, the IoT devices generate a stream of tasks denoted by $f_i$, where $i \epsilon \{1, \dots, I\}$. A task $f_i$ is characterized by the tuple $(G_i, c_i, s_i, s'_i, D_i)$, representing its arrival time $G_i$, computational demand $c_i$ (in cycles), its associated size $s_i$ (in bits), the size $s'_i$ of the task's output after the task has been processed, and the deadline $D_i$ (in seconds) by which the task must be executed. When a device offloads a task to an edge or cloud server, the computing unit must transmit the task's output back to the originating device. The size of the output task is defined as $s'_i = \rho \cdot s_i$, where $0 < \rho < 1$ denotes the fraction of the original task size. Computing units, including local processors of IoT devices ($J$), edge node ($Q$), and cloud server (C) are characterized by their processing capabilities $\chi_m$ (in cycles/second), $m \epsilon \{J, Q, C\}$. The processing time for task $f_i$ on a selected computing unit $m$ can then be expressed as $t_p^i = \frac{c_i}{\chi_m}$.

We assume an FDD-OFDMA (Frequency Division Duplexing - Orthogonal Frequency Division Multiple Access) cellular network. The amount of data that can be transmitted within a radio resource block (RB) $\kappa \epsilon K$ is:

$$R_\kappa = \tau_\kappa . BW_\kappa \times \log_2(1 + \gamma_\kappa), \quad (1)$$

where $\tau_\kappa$ represents the duration of the RB, $BW_\kappa$ denotes the RB's bandwidth, and $\gamma_\kappa$ is the instantaneous Signal-to-Interference-plus-Noise Ratio (SINR). To capture channel fading effects, we adopt a Rayleigh fading channel model. As a result, the instantaneous SINR $\gamma_\kappa$ follows an exponential distribution with mean $\bar{\gamma}_\kappa$. Considering $K_i^{UL}$ as the total number of RBs scheduled for the uplink transmission of task $f_i$ when it is offloaded, we can calculate the total amount of data transmitted in the uplink as $R_i^{UL} = \sum_{\kappa \epsilon K_i^{UL}} R_\kappa$. Similarly, the total amount of data transmitted in the downlink is computed as $R_i^{DL} = \sum_{\kappa \epsilon K_i^{DL}} R_\kappa$, where $K_i^{DL}$ represents the RBs scheduled for the downlink transmission of task $f_i$.

The execution time $T_i$ is defined as the time elapsed from the arrival time $G_i$ of task $f_i$ until the end of its execution. It consists of the processing time ($t_p^i$) spent at the computing unit, and the communication latency incurred when the task is offloaded to the edge or the cloud. In this case, the communication latency comprises the uplink transmission time ($t_{UL}^i$) required to send the task and the downlink transmission time ($t_{DL}^i$) required to receive the task output at the originating IoT device. A task is considered successfully executed if its execution time $T_i$ does not exceed its deadline $D_i$.

## III. DETERMINISTIC TASK OFFLOADING STRATEGIES

### A. *Deterministic task offloading*

The authors proposed in [7] a deterministic task offloading strategy aimed at maximizing the number of tasks executed within their deadlines (i.e., tasks for which $T_i \leqslant D_i$). To this aim, the strategy leverages the diversity of task deadlines to decide where tasks should be offloaded and processed across the continuum. The objective function is defined to minimize the number of tasks that miss their deadlines as:

$$\min_{a_{i,m}} \sum_{i \epsilon I} Y\left(\frac{T_i(a_{i,m})}{D_i}\right), \quad (2)$$

where the action space $a_{i,m} = [a_{i,J}, a_{i,Q}, a_{i,C}]$ is a vector in which only one element can be equal to 1, representing the selected computing unit $m \epsilon \{J, Q, C\}$. $T_i(a_{i,m})$ reflects that the selected computing unit determines the execution time $T_i$ experienced by the task $f_i$. $Y(x)$, with $x = \frac{T_i}{D_i}$, is a penalty function defined as:

$$\Upsilon(x) = \begin{cases} 0, & 0 \leq x \leq 1, \\ \Omega, & x > 1, \end{cases} \quad (3)$$

that assigns a high penalty value $\Omega$ to tasks that are not executed in time (i.e., $T_i > D_i$); and no penalty to those executed within their deadlines (i.e., $T_i \leq D_i$) regardless of their exact execution time $T_i$. As a result, the strategy does not differentiate between tasks executed within time, nor does it aim to reduce $T_i$. Instead, it focuses on maximizing the number of tasks executed within their deadlines.

### B. *Deterministic task offloading with execution timing control*

Like most existing task offloading strategies, the deterministic task offloading scheme in [7] focuses on selecting where a task should be executed (locally or offloaded to the edge or cloud). This paper proposes introducing a timing control process into the deterministic task offloading strategy that leverages the tasks' latency budgets, i.e., from tasks' arrival to their deadlines, to jointly determine where tasks are executed and when their execution should begin. The objective is to improve resource utilization across the continuum, and consequently the number of tasks executed in time, by distributing the workload in time and space (i.e., across nodes in the continuum).

The proposed scheme extends the action space of the original deterministic task scheduling scheme by introducing the tuple $\varphi_i =< a_{i,m}, O_i >$, which captures both the spatial (where) and temporal (when) dimensions of task execution. In particular, the scheme decides for each task $f_i$ the tuple $\varphi_i =< a_i, O_i >$ where $a_{i,m}$ is the same vector as defined for the deterministic task scheduling scheme to represent the selected computing unit, and $O_i$ represents a controlled offset added to the arrival time of the task $f_i$. This offset then determines when the task starts being processed if it is handled locally at the IoT devices, or when its offload to the edge or cloud starts. Thus, the execution time $T_i$ in the proposed scheme includes this offset and is defined as $T_i = O_i + t_p^i + t_{UL}^i + t_{DL}^i$. To evaluate the optimality of a chosen action space $\varphi_i$ for a task $f_i$, the proposed scheme considers its execution time (i.e., $T_i(\varphi_i)$), carefully balancing the added offset $O_i$, the communication latency ($t_{UL}^i + t_{DL}^i$) and the processing time $t_p^i$, to ensure that it does not exceed the task's deadline $D_i$. Then, the proposed scheme tries to minimize the following objective function:

$$\min_{\varphi_i} \sum_{i\epsilon I} \Upsilon\left(\frac{T_i(\varphi_i)}{D_i}\right), \tag{4}$$

where $\Upsilon(x)$ is defined as in (3) with $x = \frac{T_i(\varphi_i)}{D_i}$. This penalty function prioritizes deadline satisfaction while optimizing the allocation tuple $\varphi_i$ and imposes no penalty if the execution time is within the task's deadline ($T_i \le D_i$). Conversely, a heavy penalty $\Omega$ is applied when the execution time leads to a deadline violation. This definition ensures that the scheme controls the tasks' execution times across the continuum without missing task deadlines.

### C. Constraints

We consider the following constraints for the optimization problems defined in (2) and (4), with the distinction that $O_i = 0$ for all tasks in the optimization problem in (2):

$C_1: a_{i,m} \epsilon \{0,1\}, \forall m \epsilon \{J, Q, C\}, \forall i \epsilon I,$

$C_2: \sum_m a_{i,m} \le 1, \quad \forall i \epsilon I,$

$C_3: \sum_{i\epsilon I} c_i a_{i,m} \le X_m, \qquad \forall m \epsilon \{J, Q, C\},$

$C_4: \sum_{i\epsilon I} b_{i,\kappa}^{UL//DL} = 1, \forall \kappa \epsilon \mathrm{K}^{UL//DL},$

$C_5: s_i \le \sum_{\kappa \epsilon K_i^{UL}} R_\kappa, \forall i \epsilon \mathrm{I},$

$C_6: s'_i \le \sum_{\kappa \epsilon K_i^{DL}} R_\kappa, \forall i \epsilon \mathrm{I},$

$C_7: b_{i,\kappa}^{UL} = 0, if\ t(\kappa) < G_i + O_i, \forall i \epsilon \mathrm{I},$

$C_8: \sum_m a_{i,\mathrm{m}} = 0,\ \ if\ t < G_i + O_i + t_{UL}^i, \forall i \epsilon \mathrm{I},$

$C_9: b_{i,\kappa}^{DL} = 0, if\ t(\kappa) < G_i + O_i + t_p^i + t_{UL}^i, \forall i \epsilon \mathrm{I},$

$C_{10}: O_i < D_i, \forall i \epsilon I.$

Constraints $C_1$-$C_2$ enforce that each task is allocated to a single computing unit. Constraint $C_3$ ensures that the total computational demand of tasks assigned to a computing unit does not exceed the maximum processing capacity of that unit. Binary variables $b_{i,\kappa}^{UL}$ and $b_{i,\kappa}^{DL}$ indicate whether communication resource RB $\kappa \epsilon K$ is assigned to task $f_i$ for offloading to the edge or cloud and returning the task's output, respectively, and constraint $C_4$ ensures that each RB is allocated to at most one task. Constraints $C_5$ and $C_6$ guarantee that the total channel data capacity of the allocated RBs is sufficient to transmit the task in the uplink and its output in the downlink, respectively. Constraints $C_7 - C_9$ enforce the temporal ordering of operations: uplink transmission can only occur after the task has been generated and the established offset (if any) has elapsed; task processing should not start until the uplink transmission is completed; and downlink transmission of the task's output should not start until task processing has been completed. Finally, constraint $C_{10}$ ensures that the offset value for each task is constrained to be less than its deadline.

### D. Genetic Algorithm-Based Solution

The optimization problems in (2) and (4) are NP-hard, and we apply a genetic algorithm to obtain near-optimal solutions [9]. We define a near-optimal solution as one where the objective function reliably converges, with improvements plateauing across successive iterations. The specific algorithm parameters, including a population of 1,000 individuals run for 20 generations, with 25% elitism crossover and 20% mutation, were determined empirically through preliminary testing. This parameter selection ensures sufficient solution diversity [10] and avoids premature convergence; in our tests, the fitness values consistently stabilized before the 20-generation limit, demonstrating that these values are sufficient to achieve convergence.

## IV. Performance Evaluation

### A. Evaluation Scenario

We compare the performance of the proposed deterministic task offloading scheme with execution timing control against the original deterministic task offloading scheme that only decides where to offload the task. These will be referred to as *Proposal* and *Deterministic*, respectively. For both schemes, the penalty parameter $\Omega$ of their penalty functions $\Upsilon(x)$ is set to a value of 100. We also consider a common reference task offloading approach [3]–[5] as a benchmark. This will be referred to as *Min-Latency*. In this scheme, the action space is limited to selecting the computing unit (i.e., $a_{i,m}$), as in the original deterministic task offloading scheme, but with the primary objective of minimizing the execution time $T_i$ of each individual task $f_i$. Consequently, its objective function is defined as:

$$\min_{a_{i,m}} \sum_{i\epsilon I} T_i(a_{i,m}) \,. \tag{5}$$

To ensure a fair comparison with deterministic task offloading strategies, this benchmark scheme also considers the constraints $C_1$-$C_{10}$ defined in Section III.C, with $O_i = 0$ for all tasks, as tasks are executed immediately without any offset to start the local processing or offloading to the edge or cloud. The optimization problem in (5) is also solved using the genetic algorithm presented in Section III.D.

Without loss of generality, we consider a connected mobility scenario with six IoT devices (four smartphones and two vehicles) that can locally execute their tasks or offload them to the edge or cloud. This study does not focus on a specific IoT application. Instead, the characteristics of the generated tasks are modeled to reflect diverse computing demands and sizes. In our simulations, the aggregate task arrival rate across all devices varies from 0 to 500 tasks per second. The size of each task ($s_i$) is uniformly distributed within the range (3.2-6.4) Mbits, while the required computing demand ($c_i$) is modeled following a uniform random distribution within the range (50-100) Mcycles per task [11]. This results in an aggregate computational demand of up to 50 Gcycles/s. Based on commercial off-the-shelf specifications, we consider local processing capabilities ($\chi_J$) of 2 GHz (or Gcycles/s) and 8 GHz for the smartphones and vehicles, respectively [12], which are insufficient to sustain the peak task load in the scenario. The edge node ($\chi_Q$) and cloud server ($\chi_C$) have processing capacities of 90 GHz and 150 GHz, respectively [13], which can be exploited to offload tasks from IoT devices. Following a standard processor-sharing model [14], computing units (at the IoT devices, edge, and cloud) share their processing capacity equally among the assigned tasks. Then, the actual processing time $t_p^i$ required for a task $f_i$ depends on the number of tasks concurrently allocated to the computing unit. After a task has been processed, the size of the output task ($s'_i$) is set to a fixed fraction $\rho$ (set equal to 0.15) of the original task size. Task deadlines ($D_i$) are randomly selected from the discrete set of {60, 70, 80, 100} ms [3]. We consider an FDD-based OFDMA wide-area cellular network modeled following 5G NR specifications. The network operates over a total bandwidth of 50 MHz. The NR physical layer parameters include a subcarrier spacing (SCS) of 30 kHz and a slot duration of 0.5 ms. The average SINR experienced in the cellular connection between the IoT devices and the base station is set to 30 dB.

### B. *Results*

Fig. 1 illustrates the satisfaction ratio achieved by the evaluated schemes under varying task arrival rates. The satisfaction ratio is defined as the proportion of tasks executed within their deadlines. It should be noted that tasks that are not satisfied are still executed, but only after their deadlines have elapsed. This ratio reflects the average performance across different task types with deadlines randomly selected in the range [60, 70, 80, 100] ms. As expected, under low task arrival rates, all task offloading schemes meet the tasks' deadlines due to the availability of sufficient resources. However, as the task arrival rate increases, the communication and computing resources in the IoT-edge-cloud continuum become overloaded if their usage is not correctly managed, leading to a reduction in the satisfaction ratio. The *Min-Latency* approach exhibits the earliest and steepest performance degradation as the task arrival rate increases. This occurs because it manages tasks to minimize individual task execution time without considering task latency budgets. Even though this decision may be optimal for an individual task, it compromises overall task satisfaction when the task arrival rate increases, as shown in Fig. 1. The *Deterministic* scheme outperforms the Min-Latency approach by leveraging task deadline budgets to offload tasks across the continuum, enabling a higher number of tasks to be executed within their deadlines. For example, the *Deterministic* scheme's satisfaction ratio remains above 80% until the task arrival rate is 210 tasks/sec, while *Min-Latency* can only reach this satisfaction ratio until the task arrival rate is 170 tasks/sec. The proposed deterministic task offloading strategy with execution timing control demonstrates scalability, maintaining a satisfaction ratio above 80% until the task arrival rate is 330 tasks/sec. Overall, the satisfaction ratio improvement of the proposed scheme with respect to the *Min-Latency* scheme exceeds 70%. This improvement is achieved by the proposed scheme thanks to the planning and control of the tasks' execution time, which enables a more balanced distribution of tasks over time. As a result, it mitigates transient congestion at computing units and the network, allowing the continuum to accommodate and execute a larger number of tasks within their deadlines.

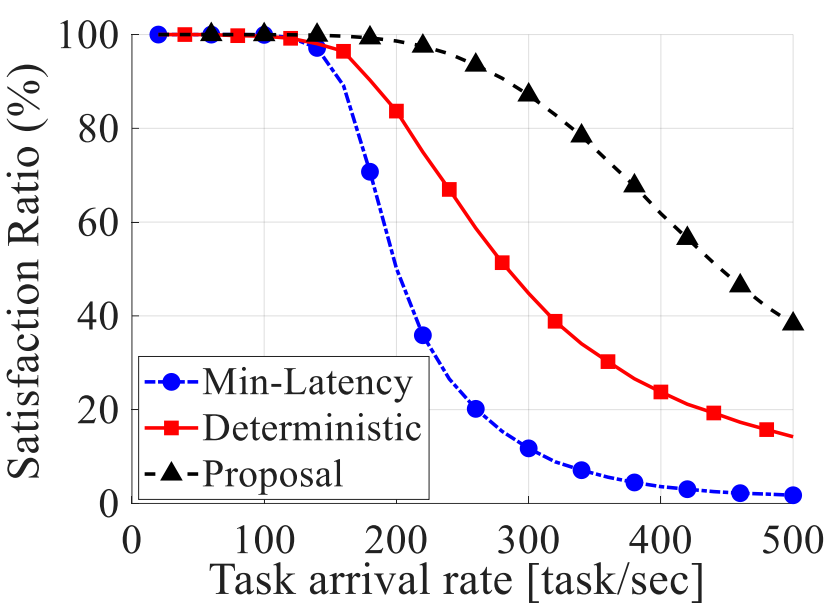

Fig. 1. Satisfaction ratio as a function of task arrival rate.

Fig. 2(a) depicts the average communication resource usage ratio as a function of the task arrival rate. The communication resource usage ratio is measured as the ratio of utilized RBs to the total number of RBs. This is computed for each generated task from the arrival time of the task until the end of its execution, and Fig. 2 reports the average value over all generated tasks. Fig. 2(a) shows that all task offloading schemes make a low usage of communication resources under low task arrival rate conditions as they rely mostly on the local processing capabilities. As the task arrival rate grows, local processors alone cannot support all tasks and the schemes have to offload tasks to the edge and cloud using the communication link. Fig. 2(a) shows that the *Deterministic* scheme starts offloading tasks at a lower task arrival rate than the *Min-Latency* scheme, even though the communication latency increases the task execution time, because this scheme aims to satisfy more tasks within the latency budget rather than minimizing the individual task execution time. Both schemes increase the communication resource usage as the task arrival rate increases, with the *Deterministic* scheme reaching its plateaus slightly below that of the *Min-Latency* scheme at high task arrival rates. In contrast, the proposed scheme consistently maintains a lower communication resource usage across all evaluated task arrival rates, keeping it below 60% (up to 40% less than *Min-Latency*) even at high task arrival rates. This is the case because the proposed scheme plans the start of the task execution considering the tasks' available latency budget. By introducing a controlled offset to the start of its execution time, the proposal allows a bigger share of tasks to be processed locally once resources become available, avoiding unnecessary offloading to the edge and cloud. For example, when the task arrival rate is 300 tasks/sec, the proposed scheme offloads only 12% of the tasks to the edge or cloud, while both *Min-Latency* and *Deterministic* offload 39% of the tasks. Fig. 2(b) complements the finding illustrated in Fig. 2(a) by analyzing the average ratio of computing units' usage across the IoT–edge–cloud continuum. The computing unit usage ratio is measured as the ratio of time that the selected computing unit is utilized from a task's arrival until the end of its execution, and Fig. 2(b) reports the average value over all generated tasks. The immediate-execution policy adopted by both the *Min-Latency* and *Deterministic* schemes leads to transient congestion of communication and computing resource utilization as shown in Fig. 2. In contrast, the proposed scheme reduces congestion (up to 15% compared to *Min-Latency*) even at high task arrival rates by performing a more balanced distribution of workload over the tasks' available latency budget.

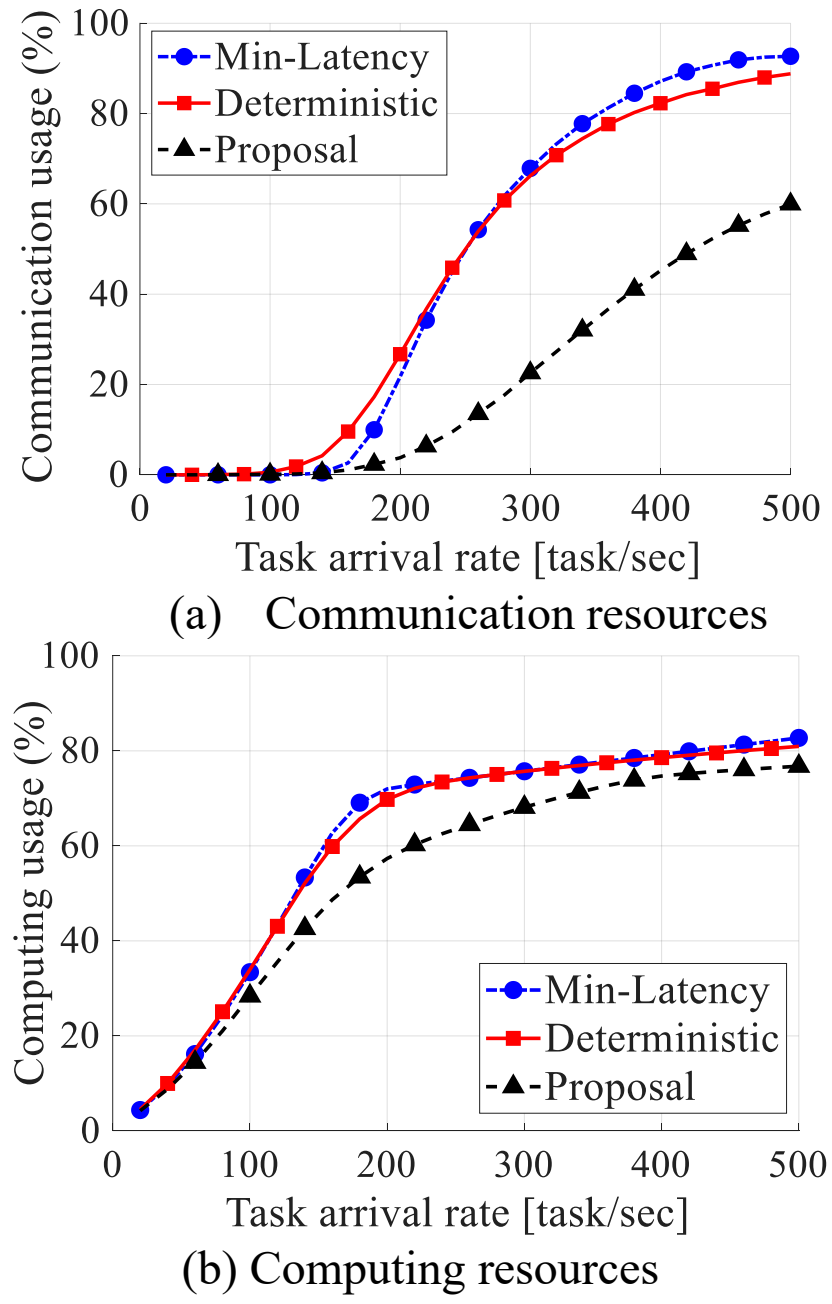

(a) Communication resources

(b) Computing resources

Fig. 2. Resources usage ratio as a function of task rate.

The superior performance of the proposed scheme in terms of satisfaction ratio (Fig. 1) and resource usage efficiency (Fig. 2), relative to *Deterministic* and *Min-Latency*, stems from the planned execution of tasks within their latency budgets. This is achieved through controlled offsets that enable tasks to start execution at appropriate times, resulting in a more balanced distribution of tasks across the continuum. As shown in Fig. 3, the proposed scheme adaptively determines an appropriate offset for each task based on its deadline (Fig. 3(a)) as well as its size (Fig. 3(b)) and processing demand (Fig. 3(c)). The results reported in Fig. 3(a) show that tasks with larger deadlines exhibit, on average, higher offsets. This occurs because the proposed scheme can leverage tasks with longer deadlines to postpone their execution further, thereby accommodating more critical tasks earlier or waiting for a decrease in local processing resources and communication resource utilization, preventing transient congestion and enabling a more distributed resource

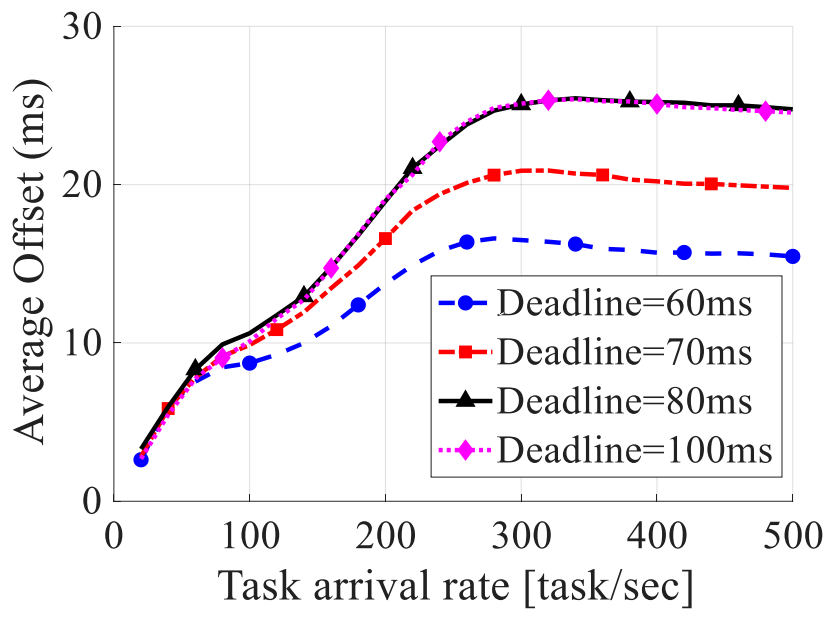

(a) Offset as a function of task deadlines for different task arrival rates

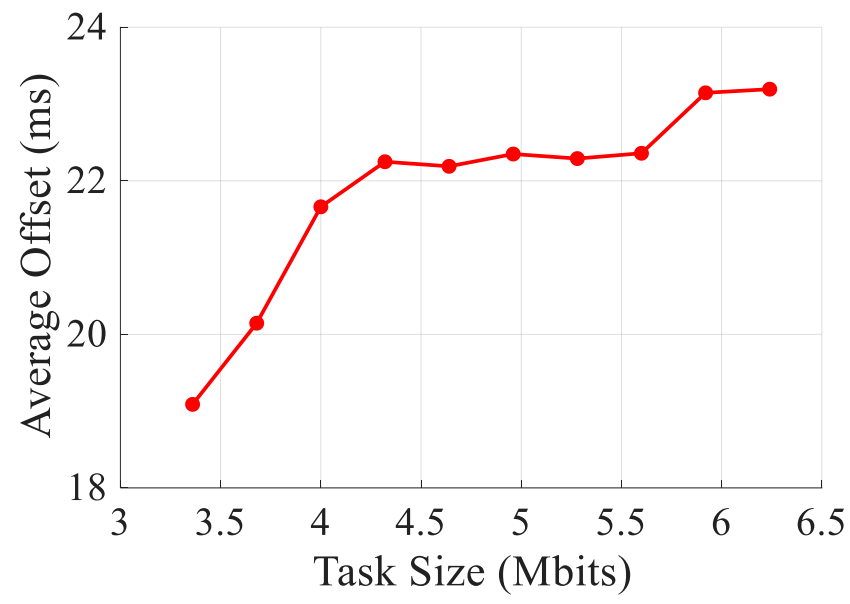

(b) Offset as a function of task size (averaged over all task arrival rates)

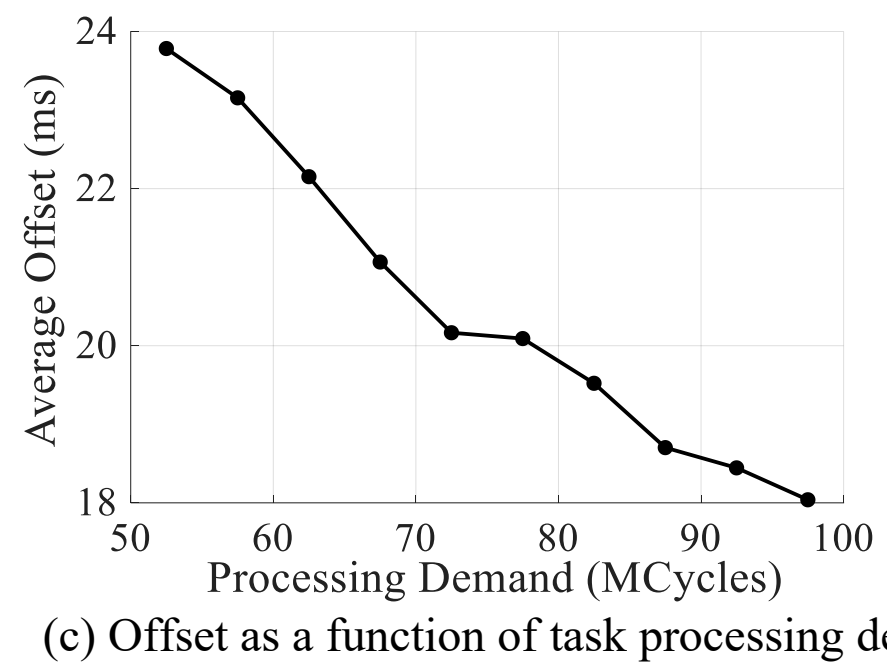

(c) Offset as a function of task processing demand (averaged over all task arrival rates)

Fig. 3. Average offset for the proposed scheme.

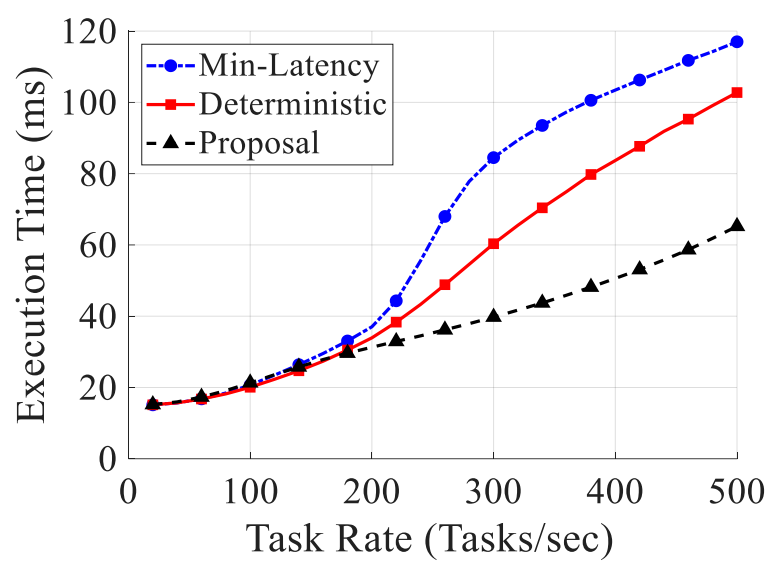

Fig. 4. Execution time versus task arrival rate.

usage (Fig. 2). Fig. 3(a) and Fig. 3(b) also show that the proposed scheme accounts for the task arrival rate and task size by increasing, on average, the offset as the task arrival rate increases, with the same objectives as in the case of increasing task deadlines, thereby enabling a more distributed execution of tasks while remaining within their deadlines. Fig. 3(c), on the other hand, indicates that tasks with higher processing demands experience, on average, lower offsets. This occurs because more demanding tasks are more frequently offloaded to the edge or cloud, where higher processing capabilities are available to complete their execution within their deadlines. For example, when the task processing demand is higher than 80 Mcycles, the proposed scheme offloads 45% of tasks to the edge or cloud, while for tasks with processing demand less than 80 Mcycles, this value is 19%. The offloading process consumes uplink and downlink communication time, which, together with their inherently higher processing times, leaves less room for postponing the start of their execution. Consequently, the proposed scheme intelligently prioritizes these tasks with high-processing demands by reducing their offset.

Finally, Fig. 4 analyzes the average execution time experienced by all generated tasks, regardless of whether they meet their deadlines (Fig. 1) or not. Fig. 4 shows that, as expected, all schemes exhibit increasing average execution time as the task arrival rate grows. However, despite the offset introduced by the proposed scheme, Fig. 4 shows it achieves lower average execution time than the *Deterministic* and *Min-Latency* schemes. The execution time reduction reaches up to 77% and 54%, respectively. These results confirm that immediate-execution task offloading policies, especially the *Min-Latency* scheme, can reduce execution time for some tasks; however, the transient congestion they may cause leads to system-level performance degradation, with many tasks missing their deadlines (Fig. 1).

## V. CONCLUSIONS

The IoT-edge-cloud continuum provides a flexible and unified framework of intelligence, communication, and computing resources that task offloading schemes can leverage to support the deterministic requirements of critical IoT applications. Most existing task offloading schemes follow an immediate-execution policy, focusing solely on selecting where tasks are processed to minimize their execution time. However, as shown in this paper, such policies can lead to transient congestion when multiple tasks coincide in time. To address this limitation, we proposed a deterministic task offloading scheme with execution timing control that jointly plans where tasks are processed and when their execution begins. The proposed scheme leverages the latency budget of tasks and distributes the workload not only across computing resources in the continuum but also over time. Our evaluation results show that incorporating execution timing control enhances the ability to support deterministic service levels. In particular, the proposed scheme reduces peak communication and computing resource congestion above 40% and 15%, respectively, leading to an improvement of over 70% in the number of tasks executed within their deadlines compared to existing benchmark schemes.